\documentclass[prb,twocolumn]{revtex4-1} 

\usepackage{graphicx} 
\usepackage{amsmath}  
\usepackage{amsfonts} 
\usepackage[normalem]{ulem}
\usepackage{color}
\usepackage{xcolor}
\usepackage{soul}
\usepackage{ulem}     
\usepackage{hyperref}

\begin{document}

\title{3D-printed microscope with illumination for undergraduate wave optics laboratory}

\author{Sergey G. Martanov}
\author{Valery A. Prudkoglyad}
\author{Arslan A. Galiullin}
\altaffiliation[Also at:]{%
HSE University, Moscow, Russia 101000
}
\author{Georgy A. Shmakov}
\altaffiliation[Also at:]{%
HSE University, Moscow, Russia 101000
} 
\author{Aleksandr Yu. Kuntsevich}
\email{alexkun@lebedev.ru} 
\affiliation{P.N. Lebedev Physical Institute of the Rusiian Academy of Sciences, Leninsky Prospekt 53, Moscow, 119991, Russia}

\date{\today}

\begin{abstract}
We present an educational tool, a  microscope with a video camera, that can be fabricated either from a standard microscope or assembled from inexpensive, commercially available components (objectives, beam splitters, LEDs, linear stages) and 3D-printed elements. Usage of interference filters in combination with white light-emitting diode (LED) illumination enables the quantitative study of optical phenomena such as refraction, interference (e.g., Newton’s rings), Fresnel and Fraunhofer diffraction. Thus, we propose an instrument that can be used to illustrate the theoretical foundations of an undergraduate optics course and beyond.
\end{abstract}


\maketitle 

\section{Introduction} 

A course of laboratories in the undergraduate optical education usually consists of separate experiments on geometrical optics, interference, diffraction, polarization, light scattering and spectroscopy.
These experiments are typically performed with light propagating in the horizontal plane, using discrete optical elements (light sources, mirrors, lenses, slits, gratings, polarizers, wave plates, etc.) installed on an optical bench or table. A similar horizontal setup is also standard in most scientific optical experiments.

In contrast, one of the fundamental optical instruments, the microscope, typically operates with vertical light propagation, which makes it particularly suitable for studying small samples. It is well known that the first Leeuwenhoek microscope was just a single lens. Two-lens schemes are described even in the introductory level physics textbooks (see e.g. Sec. 34.8 of Ref.~[\onlinecite{Sears}]). High-resolution microscopy requires the correction of optical aberrations within the objective lens, a topic that is rarely addressed in undergraduate courses. If the aberration issue is solved, objective behaves as an ideal lens and it becomes unnecessary for the user or student to understand the detailed optical layout. Commercial metallographic microscopes are, of course, powerful tools for exploration various optical phenomena: they can form images to the eyepiece or camera sensor with changeable and sub-micrometer resolution, include color and polarization filters, and support multiple illumination modes (transmitted, reflected, and epi). However, they are expensive and therefore not well suited for illustrating the undergraduate optics curricula.

In recent years, 3D printing has enabled the development of low-cost and customizable microscope systems\cite{ReviewOpenHardwMicrosc2022,GuideMicrosc3Dprinting, LowCost}.In this work, we combine microscopy and 3D printing to propose practical exercises in wave optics.

We present a vertical-microscope approach for undergraduate optics laboratories, with a special focus on the illumination system. The instrument is assembled from inexpensive and readily available components: an objective lens, LED, video camera, beam splitter, and linear stages mounted in a 3D-printed frame. The same design principles and illumination schemes also allow to upgrade commercial biological or metallographic microscopes.

Student training and interaction with the microscope begins with its assembly. During this process, students naturally encounter a number of fundamental questions: How is an optical image formed? Why is an objective lens necessary? What determines the resolution of a microscope? What is the role of numerical aperture? How should distances be calibrated, and how should the specimen be properly illuminated? Addressing these questions enables students to gain a deeper understanding of the fundamental principles of geometrical optics.

After assembly, the microscope enables direct observation of a wide range of physical phenomena. The shallow depth of focus of high-numerical-aperture (NA) objectives allows for measurements of the refractive index. The implemented illumination scheme makes it possible to perform quantitative studies of light interference phenomena, such as Newton’s rings. A lithographic mask with slits, gratings, and pinholes enables numerous practical exercises on both Fresnel and Fraunhofer diffraction using different illumination schemes. Additive manufacturing enables further extensions and student projects related to spectroscopy, polarization microscopy, lithography, two-dimensional materials, biology, machine vision and related topics. 

\begin{figure*}[!htbp]
\centering
\includegraphics[width=\textwidth]{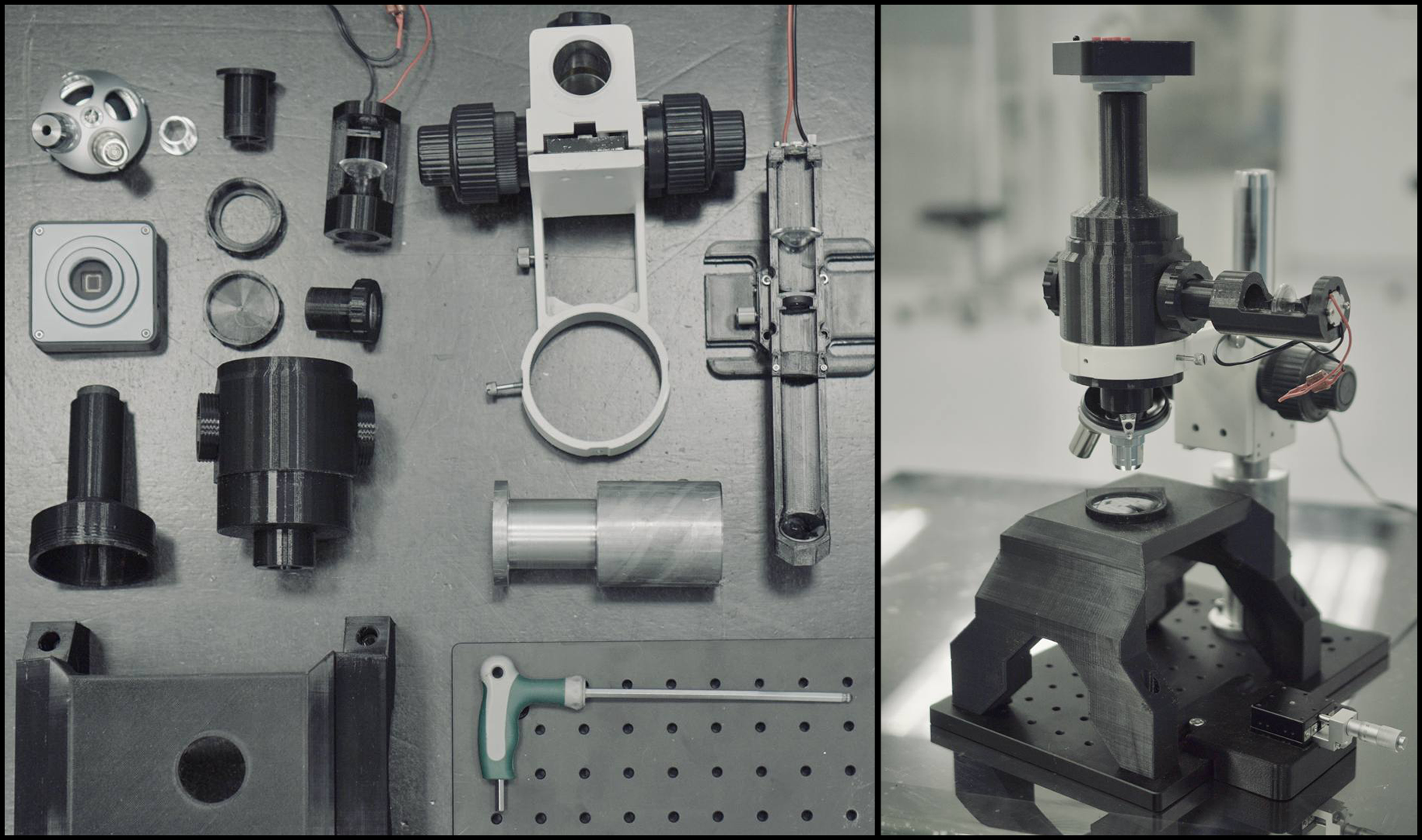}
\caption{Left: components of 3D-printed microscope. Right: microscope.}
\label{Micro}
\end{figure*}

\begin{figure*}[!htbp]
\centering
\includegraphics[width=\textwidth]{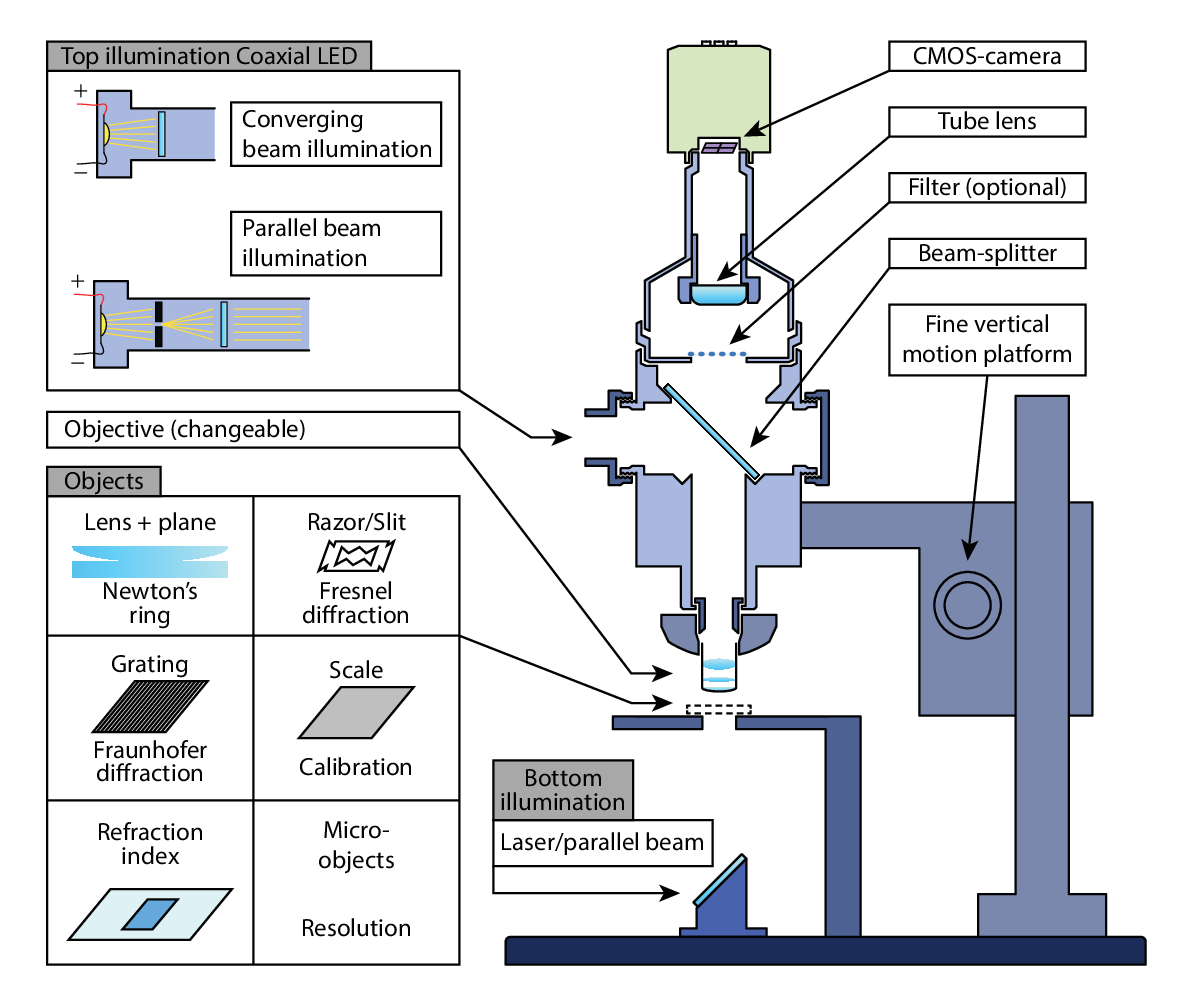}
\caption{A schematic representation of the microscope toolkit. Changeable elements are shown by dashed frames.}
\label{scheme}
\end{figure*}

\section{Microscope construction}
The microscope components and assembly are shown in Fig. \ref{Micro}. An aluminum alloy plate serves as the base. Due to its light weight, the microscope, contrary to most of metallographic and biological microscopes can be mounted on a vertical platform with a fine scale and moved as a whole. This platform is a commercial linear Z-stage with a micrometer scale and fine/coarse adjustment knobs. This stage is fixed to a 25 mm diameter vertical rod that is rigidly connected to the plate. The microscope body is 3D-printed from PETG plastic, and all mechanical connections are made using plastic threads. Coarse threads (4 mm step) are printed; fine threads are cut with specially purchased dies and taps. A standard microscope camera (\autoref{CostConsiderations}) is attached to a C-mount tube (1''/32 thread). The position of the tube lens is adjusted using the same thread. The object table is stationary and secured to the aluminum plate with M6 screws. The camera sensor is equipped with an achromatic tube lens that can be precisely focused using a fine-pitch thread.

The most crucial element of any high-performance microscope is the objective lens, a sequence of lenses designed to correct optical aberrations \cite{objectives1}.
Development of industrial lens technology made the price of reasonably high performance objective lenses affordable. The use of the standard RMS (Royal Microscope Society) thread (0.8 in – 36 tpi) allows objectives from different suppliers to be interchanged.

One must choose an appropriate objective system from the available types. For example, Nikon and many other manufacturers produce finite-conjugate objectives designed for a mechanical tube length of 160 mm (indicated as “160” on the objective barrel). Such objectives form an image directly at a distance of 160 mm from the entrance pupil and therefore do not require additional optics. However, they are not ideal for educational use, since it is difficult to maintain the 160 mm spacing with sufficient precision, and any additional element placed between the objective and the camera sensor, such as a polarizer or a filter, introduces aberrations. Moreover, standard 160 mm objectives are designed for 1 `` eyepieces, and a typical low-cost camera sensor covers only a small portion of the available field of view.

Other important objective characteristics include the immersion medium  (marked "Oil" or "W" for water) and/or cover glass correction (typically denoted as "0.17", corresponding to the standard cover glass thickness of 0.17 mm). These options increase the numerical aperture, i.e. resolution. However, they are not required for educational purposes, where it is crucial to move the object plane vertically. Objectives for epi-illumination (marked as DF - dark field) or differential interference contrast (DIC) are designed for sophisticated illumination schemes and are also excessive.

We use a so-called infinity-corrected plan objective (marked "$\infty$" on the objective barrel), in which the object should be located at the effective focal plane. The objective converts each point of the object into a parallel light beam with a corresponding inclination angle. Thus the specimen is optically projected to infinity. A key advantage of this optical scheme is that the image formation is independent of the distance between the microscope objective and the camera. This allows flat optical elements, such as beam splitters, filters, apertures or polarizers, to be inserted into the collimated beam path without affecting the image quality. The infinity-corrected configuration is therefore the most popular for scientific research.

We recommend using plan-achromatic, infinity-corrected objectives with {long} working distances, such as 5x/NA0.15, 10x/NA0.30
and 20x/NA 0.40. A longer working distance generally implies a higher cost and a lower NA, yet it greatly reduces the risk of damaging the objective by inexperienced users.
To refocus the beam from the infinity onto a photo-sensitive matrix, a secondary focusing lens, known as the tube lens, is required. Therefore a camera must be equipped with a tube lens capable of focusing at infinity, as shown in Fig.~\ref{scheme}.

Illumination in our microscope can be provided either through the objective lens (reflection scheme) or transmitted through a transparent or semi-transparent specimen (transmission scheme typical of a biological microscope), as shown in Fig.~\ref{scheme}. For top illumination, a 45$^\circ$ beam-splitting plate is placed between the camera and the microscope objective. We employ 50/50 beam splitter with a single-sided anti-reflection (AR) coating. For possible extensions, such as photoluminescence or interference microscopy this beam splitter can be replaced with a dichroic plate or a cube beam splitter, respectively. A key educational feature of our design is the 3D printed illumination system based on white LED, which can be configured for both reflection (top) and transmission (bottom) lighting.

The moving stage and optical elements such as the camera, tube lens, beam-splitter plate and objective lens are readily available from commercial platforms such as AliExpress, eBay, or Amazon. All remaining mechanical parts can be fabricated using 3D printing. 3D-printed components can also be used to upgrade existing microscopes to comparable functionality. The open-source software ToupView is compatible with most cameras and enables quantitative image analysis.

\section{Basic microscope optics}
\subsection{Infinity-corrected optical scheme}
As described above, the infinity-corrected optical scheme consists essentially of two components: (i) a tube lens with {the} camera located at its focal plane and (ii) an objective lens with the sample {placed} at the effective focal plane.To properly position the achromatic doublet tube lens, one should focus on a distant object (for example, a tree outside the window). Adjusting the lens position using the threaded mount serves as the first exercise during the microscope assembly.

Most commercial infinity-corrected objectives are designed to work with a tube lens of 200 mm focal length and an eyepiece pupil approximately 1 inch in diameter, depending on a particular supplier.
Therefore, the maximum field-of-view dimension is about 25 mm$/M$, where $M$ is the objective magnification, and the characteristic angular field of view provided by a standard micro-objective is approximately 25/200$\approx$ 0.1 rad.

The camera sensor, with a resolution of, say 3840x2106 pixels and a diagonal of about 1 cm, requires a tube lens with a focal length $\approx$100 mm to match the objective's field of view to the smaller camera sensor. Diffraction gratings ( \autoref{FraunhoferSection}) are used to calibrate the lateral resolution in pixel per millimeter for each objective.

\subsection{Illumination light source}
\label{sec:light_source}

The spectral purity of the illumination source plays a crucial role in demonstrating and quantitatively analyzing the wave-optics phenomena. An obvious choice for such sources is a laser pointer. However, laser illumination is  unsuitable for most of exercises due to speckles, i.e. inevitable spatial nonuniformities of the wave field. In contrast, incoherent LED light filtered through an interference filter provides uniform illumination with reasonable monochromaticity and performs excellently in demonstrations of wave-optics effects. The filters allow selection of a wavelength range of approximately 10-40 nm from the spectrum of intense white light.

\subsection{Top illumination scheme}

{The scheme of the top illumination, together with its photo and 3D model, is shown in Fig.} \ref{ParallelBeam} {a,b,and c, respectively. The LED light passes through a color filter, Lens 1 and} a beam splitting plate (BSP) {with} 45$^\circ$ orientation. {The BSP} has multilayer coatings on both sides: an AR coating on one side and an achromatic fixed reflection/transmission ($R/T$) ratio coating on the other.
{The focus of Lens 1 is adjusted so that the image of the LED is formed at the entrance pupil of the micro-objective, ensuring maximum intensity and a relatively uniform illumination of the field of view. The angle of Lens 1 observation should exceed 0.1 radian ($\approx 5^\circ$) to illuminate the entire field of view.  As shown in Fig.} \ref{ParallelBeam} b,c, Lens 1 is secured by an adjustable holder that enables precise alignment along the optical axis.
Regarding the beam splitter, it is instructive to discuss its operation. The operation of multi-layer dielectric coatings relies on interference and Fresnel formulas. It is instructive to ask the students to find each side. The other interesting question is which side should be on the top? The correct answer is that the AR-coated side should be on top, because coatings are never perfect and illumination should primarily be reflected from the lower side. Another useful discussion point concerns the optimal reflection/transmission ratio. Usually the problem at high magnifications is low amount of light. Since illumination is proportional to reflectivity ($R$) and the detected optical signal is proportional to transmittance ($T$), for lossless dielectric coatings where $T = 1 - R$, the product $TR$ is maximized when $R = T = 0.5$.

\subsection{Bottom illumination scheme}

In standard biological microscopes, bottom illumination is usually provided by light passing through a condenser lens, which produces a wide range of angles and allows full utilization of the objective’s numerical aperture. Although it is possible to integrate a condenser into our scheme, large illumination angles are unnecessary for wave-optics phenomena, as they tend to destroy interference patterns.
We therefore use collimated light for diffraction experiments and slightly diverging light for bottom-illumination geometry of the Newtons rings. 

\begin{figure*}[!htbp]
\centering
\includegraphics[width=\textwidth]{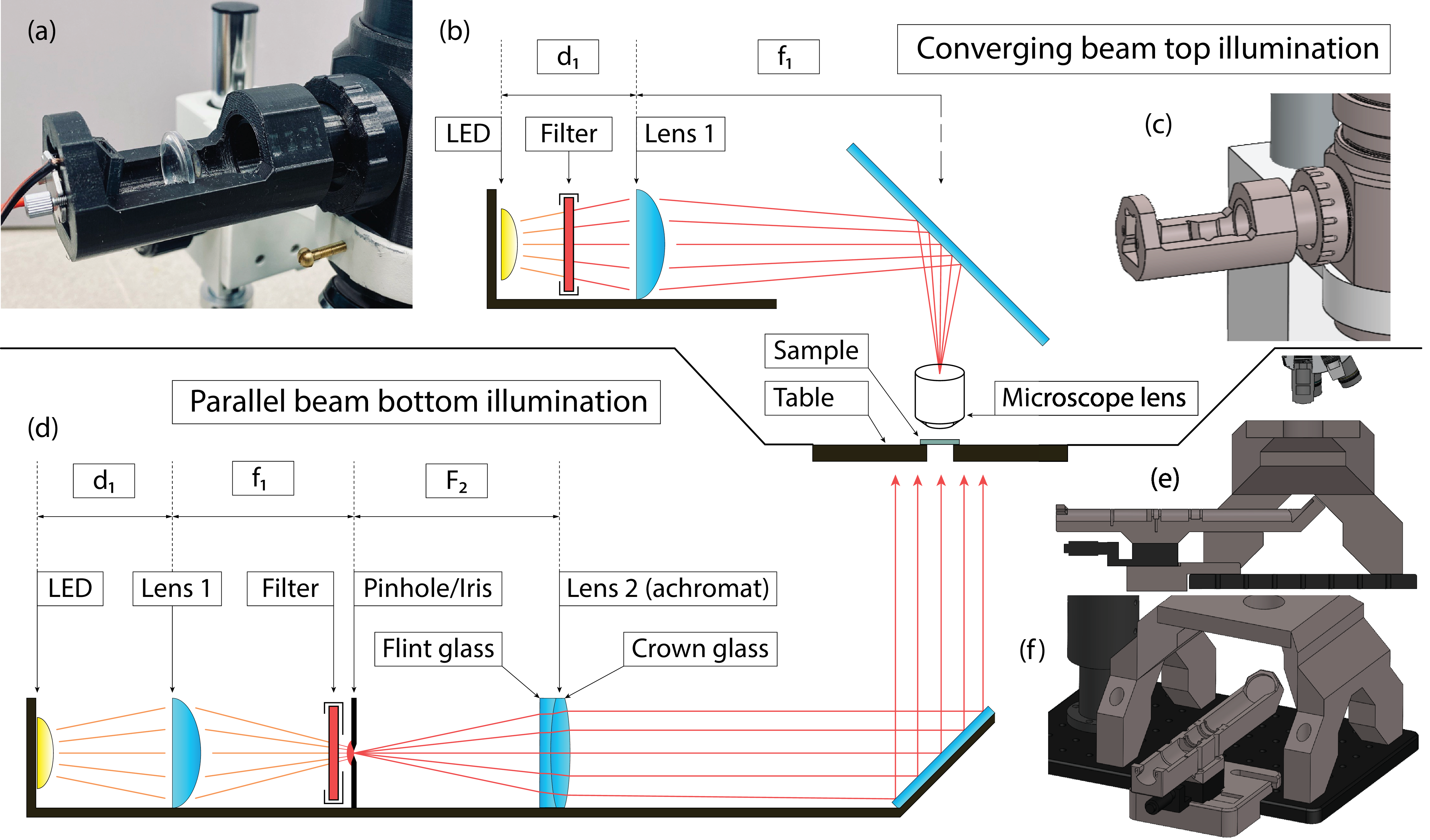}
\caption{(a) {Schematics of the top-illumination scheme. Photo(b) and 3D model(c) of top illumination. (d)} Schematics of the collimated light source {for nearly parallel light illumination}. The positions of Lenses 1 and 2 are determined using the thin lens equation. { 3D views of parallel bottom illumination a shown in panels (e) and (f)}}
\label{ParallelBeam}
\end{figure*}

To switch between parallel and slightly divergent light we use a monochromatic collimator. The optical scheme and 3D models are shown in Figs. \ref{ParallelBeam} d,e and f, respectively. The central element of the design is an iris diaphragm that acts as a pinhole, i.e point source when it is almost closed. The diaphragm is positioned within a focal length distance in front of Lens 2 to  collimate the emerging beam. We deliberately chose Lens 2 to be achromatic so that its focal length remains unchanged when the emission wavelength is altered by replacing the interference filter.

Lens 1 is positioned between the LED and the pinhole to maximize light collection and ensure precise focusing onto the pinhole. The positions of both lenses are determined from the thin-lens equation.

\subsection{Microscope calibration and objective parameters}
The theoretical resolution limit of an optical microscope is governed by diffraction and is approximately given by:
\begin{equation}
\label{eq:resolution}
    \delta \approx \frac{\lambda}{2\,\mathrm{NA}}    
\end{equation}
where $\lambda$ is the wavelength of light {and $NA$ is the numerical aperture of the objective.}
 In our setup, a 4K camera has pixels with a geometrical size of about {3~}$\mu$m. To estimate the effective object-space resolution, this pixel size must be divided by $M/2$, where $M$ is the nominal magnification of the objective, and the factor of 2 accounts for the use of a 100 mm tube lens instead of the standard 200 mm. For all objectives used in this study, the resulting pixel-limited resolution remains several times finer than the diffraction limit.

A set of microstructured elements can be used to check the diffraction limit.
For example CD disc has a period of 1.6 $\mu$m, DVD disc has {a} period $0.74$ $\mu$m, HD-DVD and Blu-ray {discs} have periods {of }0.4 and 0.32  $\mu$m, respectively.
The latter periods could only be resolved with violet illumination and high-aperture objectives.

A useful exercise for students is to determine the vertical translation stage scale. By sequentially focusing on the top and bottom surfaces of a calibrated brick ( $h_2$ and $h_3$ in Fig.~\ref{refraction1}) and counting the number of knob rotations required to shift focus between them, one can calibrate the vertical scale.

\section{Microscope operation. Exercises.}

\subsection{Refraction index measurement}
\label{Refr}
\begin{figure}[!htbp]
    \centering
    \includegraphics[width=0.5\linewidth]{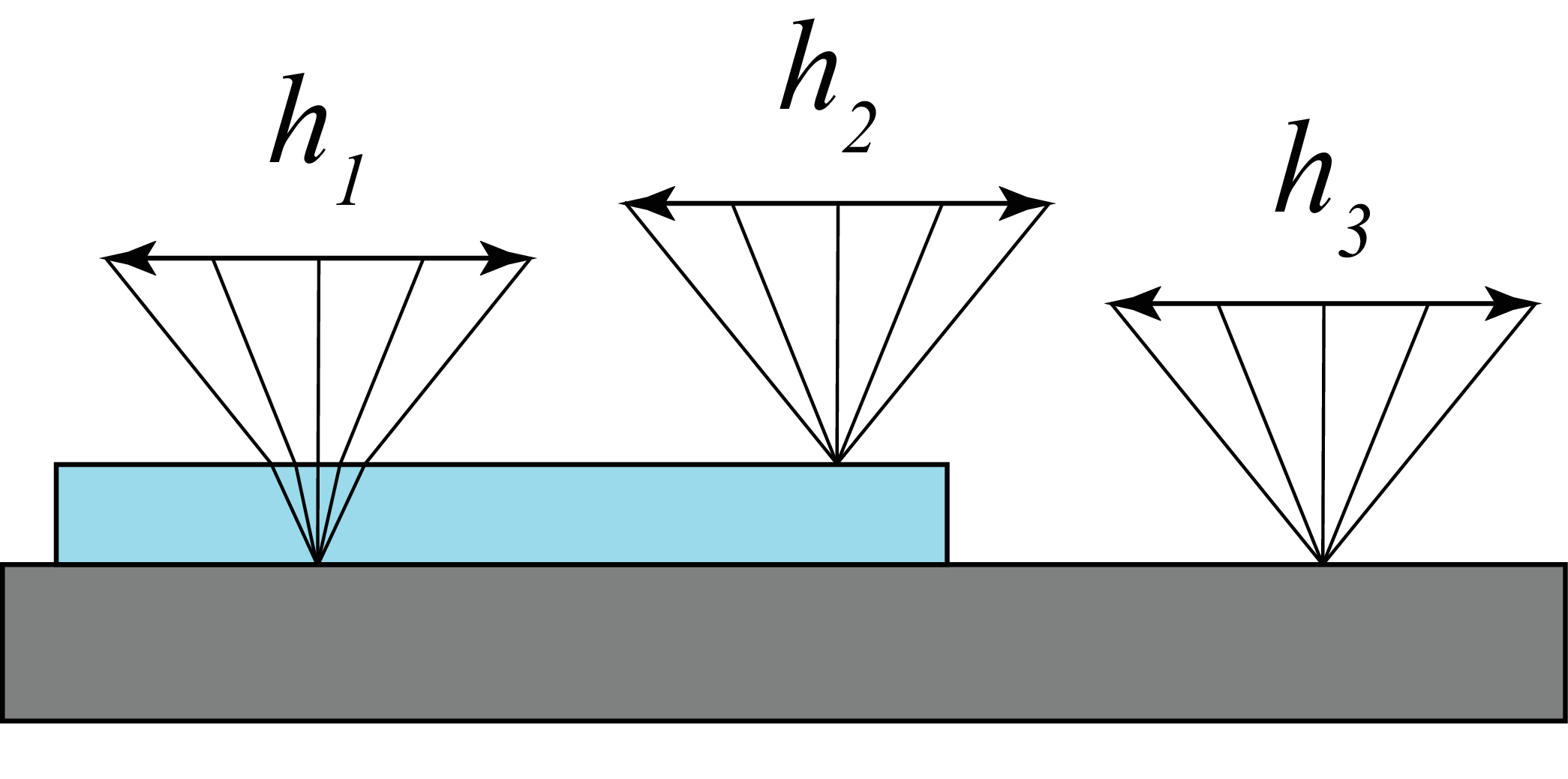}
    \caption{Schematics of the experiment on refraction coefficient determination}
    \label{refraction1}
\end{figure}

High aperture objectives have rather small depth of focus about $\lambda/NA^2$, i.e. vertical distance within which the image remains sharp. This property enables measurements of the focal plane displacement with micrometer resolution and, hence determine the optical density of the transparent parallel plates.
The geometry of the experiment is presented in Fig.\ref{refraction1}, and the refraction coefficient is given by the formula:
\begin{equation}
\label{refraction}
n=\frac{h_2-h_3}{h_2-h_1},
\end{equation}

where $h_1, h_2$ and $h_3$ are the vertical positions of the objective when focused on the bottom surface of the transparent plate, the top surface of the plate, and the substrate, respectively. For 1 mm thick plate we measured  refraction indices of glass, sapphire with $\sim$1\% precision.

\subsection{Newton`s rings}
Newton`s rings are one of the most spectacular manifestations of interference, typically observed between the plane glass and the convex lens, see {e.g.} Refs.\cite{NewtonHeNe,giantrings}.

\begin{figure*}[!htbp]
    \centering
    \includegraphics[width=\textwidth]{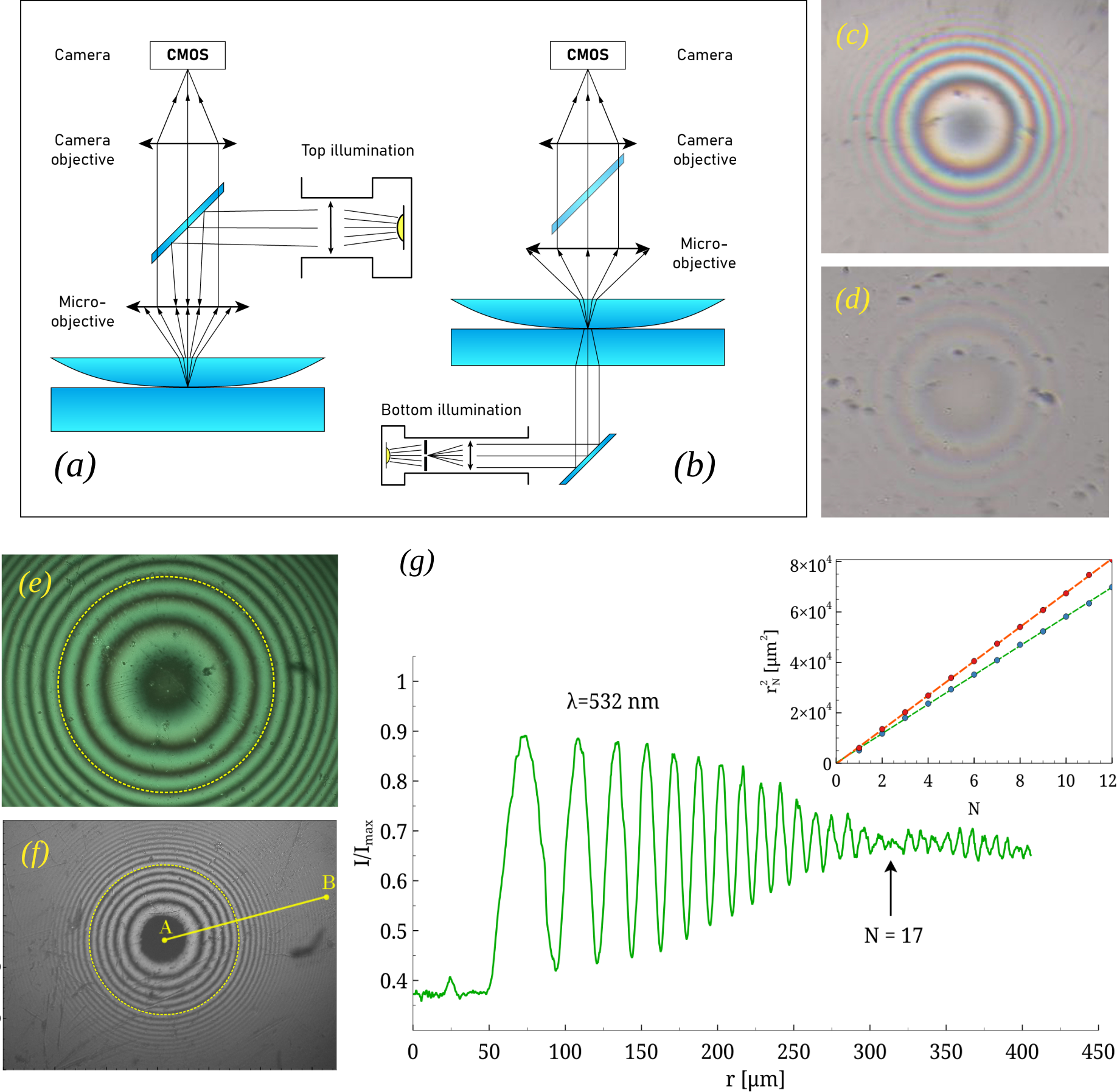}
    \caption{Schematics of the top (a) and bottom (b) geometry for the observation of Newton rings. Newton’s rings pattern under white-light illumination for both geometries shown on panels (c) and (d). Panels (e) and (f) show ring pattern for filtered white LED light ($\lambda\approx532\,$nm) for lenses of different curvatures (dashed circles are shown as a guide to the eye). Panel (f) shows RGB-averaged and normalized intensity instead of direct camera image. Panel (g) show radial dependence of intensity along A-B line on panel (f) (shown in yellow) and the inset show the squared maxima radii as function of the maxima number for two wavelengths $\lambda_1\approx 650\,$nm (higher slope), $\lambda_2\approx 532\,$nm (lower slope) and lens with $R\approx 1\,$cm.}
    \label{Newtonrings}
\end{figure*}

There are two main schemes for observing Newton's rings: in reflected and transmitted light, shown in Figs.~\ref{Newtonrings}a and ,\ref{Newtonrings}b, respectively. In the reflection geometry, interference occurs between the wave reflected from the spherical surface of the lens and the wave that passes through the lens and is reflected from the flat glass surface. Since both waves are reflected from interfaces with similar optical properties, their amplitudes are close to each other, ensuring high interference pattern contrast. The phase difference between monochromatic waves at a distance $r$ from the optical axis of the lens is determined by the following relation\cite{ghatak2009optics, hecht5ed}
\begin{equation}
\label{eq:newton-phase}
\Delta\varphi=k\Delta+\pi,
\end{equation}
where $k=2\pi/\lambda_0$ is the vacuum wave number, $\Delta=r^2/R$ is twice the thickness of the air gap between the lens and the glass, $R$ is the radius of curvature of the lens surface. The additional term $\pi$ in {Eq.~}\ref{eq:newton-phase} {comes from difference in }
the reflection conditions for the two interfering waves: one wave reflects from a boundary with a optically dense medium, while the other reflects from a boundary with a vacuum. This leads to the following expression for the radii of the bright rings\cite{ghatak2009optics, hecht5ed}:  

\begin{equation}
\label{eq:newton-reflected}
r_m = \sqrt{\lambda_0 R \left(m - \dfrac{1}{2}\right)}, \quad m = 1, 2, \ldots      
\end{equation}
In particular, intensity minimum (a dark spot) is located at the center of the pattern.

In transmitted light (Fig.~\ref{Newtonrings}b), interference occurs between the wave that passes entirely through the glass and the lens and the wave that undergoes double reflection -- first from the spherical surface of the lens and then from the flat surface of the glass. Due to the additional reflection, no extra phase difference arises, and the formula for the radii of the bright rings changes:  

\begin{equation}
\label{eq:newton-transmitted}
r_m = \sqrt{\lambda_0 R m}, \quad m = 0, 1, 2, \ldots  
\end{equation}
Unlike the reflective geometry, the amplitudes of the interfering waves in this case differ significantly, {diminishing the }visibility of the interference pattern.

To observe the Newtons rings in both geometries one has to locate the point of contact.  The microscope should be sequentially focused on the on the bottommost part of the lens and top of the glass slide, while moving to the point of contact where distance between these two positions vanishes and the Newtons rings inevitably appear.

White LED does not allow to observe more than 5 rainbow-colored rings as seen in Fig.~\ref{Newtonrings}c,d.
A hl{nearly monochromatic} illumination allows to get many rings and perform quantitative measurements. The ring pattern for the reflected light illumination scheme (see Sec.~\ref{sec:light_source}) is shown in Fig.~\ref{Newtonrings}e,f (average wavelength 532\,nm). Fig.~\ref{Newtonrings}e displays the raw color image captured directly by the camera, while Fig.~\ref{Newtonrings}f presents the averaged intensity across three color channels for a lens with a smaller radius of curvature. Fig.~\ref{Newtonrings}g shows an example of the intensity profile of the observed pattern, derived from the image in Fig.~\ref{Newtonrings}f. The $r_m^2(m)$ dependence is nearly linear, in excellent agreement with the theoretical predictions, as the inset in Fig.~\ref{Newtonrings}{g shows for two different wavelength}.

Observation of the Newtons rings in quasi-monochromatic light also provides a vivid illustration of the coherence length concept. Fig.~\ref{Newtonrings}g clearly shows the non-monotonic decay of fringe visibility, a hallmark of quasi-monochromatic light interference. By determining the ring number $N$ where the contrast is minimal (in our case, $N=17$), the spectral width $\delta\lambda$ can be estimated using the formula\cite{ghatak2009optics}:
\begin{equation}
\label{eq:newton-coherence}
    \delta\lambda\approx\lambda/N    
\end{equation}
For this experiment the value $\delta\lambda\approx30\,$nm,  agrees well with the independently measured spectral bandwidth of the optical filter. An example of the instructional materials on Newton's rings could be found in Supplementary.

\subsection{Fresnel diffraction}
\label{FresnelSection}
iffraction is another striking manifestation of the wave nature of light. It consists of the blurring of the boundaries of the geometric shadow region formed behind an obstacle in the path of a wave.  According to Huygens, every point on a wavefront acts as a source of secondary spherical wavelets, so that the wavefront at later times can be found as the envelope of the secondary wavelets. 
Later Fresnel supplemented Huygens’ principle by stating that the wave field amplitude can be determined as the result of the interference of secondary wavelets, taking into account their phase relationships~\cite{ghatak2009optics, hecht5ed}.

When analyzing light diffraction, two distinct diffraction regions are typically distinguished based on the value of the Fresnel number $f$, defined as:
\begin{equation}
\label{eq:dif-zones}
f = \dfrac{a^2}{\lambda L},    
\end{equation}
where $a$ is the characteristic size of the obstacle in the light wave's path, $\lambda$ is the light wavelength, and $L$ is the distance from the obstacle to the observation point. The region, where $f \sim 1$, is commonly called the Fresnel diffraction region, while the far-field region, where $f \ll 1$, is referred to as the Fraunhofer diffraction region.

The microscope is very useful for observation of Fresnel diffraction, as it projects the wave field from the object plane directly onto a CMOS sensor while preserving the phase distribution. Consequently, one can move the microscope vertically and see how the pattern evolves.

We observe Fresnel diffraction from a lithographically fabricated single slit and a straight edge of an opaque mask. Even for these simplest configurations, the diffraction {has no mathematical description through elementary functions}. Nevertheless, the diffraction patterns can be qualitatively and semi-quantitatively understood within the concept of Schuster zones and the Cornu spiral\cite{ghatak2009optics}.

\begin{figure}[!htbp]
    \centering
    \includegraphics[width=0.95\linewidth]{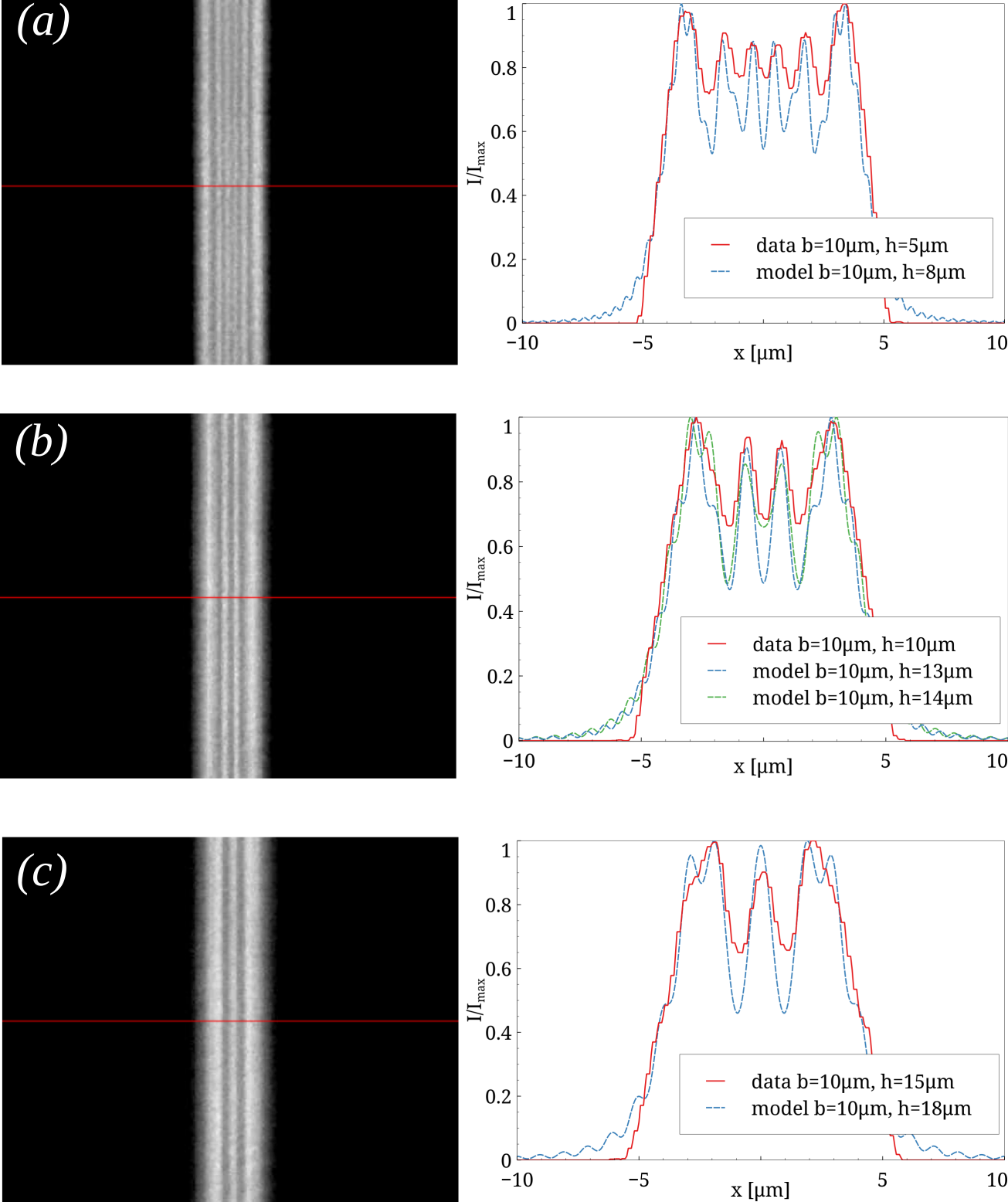}
    \caption{Fresnel diffraction patterns for the single slit (width $10\,\mu m$). Panels (a)-(c) show RGB-averaged and normalized intensity patterns captured by digital camera (left part) and corresponding intensity profiles (right part). Images were taken for microscope positions, corresponding the objective focused (a)$5\,\mu m$, (b) $10\,\mu m$, (c) $15\,\mu m$ above slit plane. Dashed curves show theoretically calculated patterns for the same distances plus about $3\,\mu m$.}
    \label{fig:fresnel}
\end{figure}

Fig.~\ref{fig:fresnel} presents the results of observing Fresnel diffraction from a single {10\,$\mu$m width} slit illuminated by green light with an average wavelength of 532\,nm. The experiment was performed in the following sequence: first, the microscope was focused on the mask plane, and then it was controllably raised to various heights above the initial position. The images in the left panels Fig.~\ref{fig:fresnel}(a, b, c) correspond to different positions of the microscope objective relative to the mask plane. The right panels show the corresponding transverse intensity profiles. The solid curves are obtained experimentally, while the dashed curves are numerically calculated diffraction patterns.

Numerical simulation of the pattern was performed based on the following formula, derived using the Fresnel diffraction integral (see e.g. Refs.~\cite{ghatak2009optics, hecht5ed}):

\begin{align}
&I_P\sim[C(u_2+u_1)-C(u_2-u_1)]^2+\nonumber\\&+[S(u_2+u_1)-S(u_2-u_1)]^2,\label{eq:fresnel-i}\\
\nonumber\\
&u_1=\dfrac{b}{2}\sqrt{\dfrac{\pi}{\lambda h}},\quad u_2=x\sqrt{\dfrac{\pi}{\lambda h}},\label{eq:fresnel-u}\\
\nonumber\\
&C(u)=\sqrt{\dfrac{2}{\pi}}\int\limits_{0}^{u}\cos\tau^2 d\tau,\quad S(u)=\sqrt{\dfrac{2}{\pi}}\int\limits_{0}^{u}\sin\tau^2d\tau\label{eq:fresnel-cs},
\end{align}
where $I_P$ is the intensity at the observation point, $b$ is the slit width, $h$ is the distance from the slit to the observation plane, and $\lambda$ is the wavelength
. {For better agreement we add up an} additional vertical displacement {$h_0\sim3\, \mu$m to} the microscope in the model relative to the physically measured height. Thus, the parameter $h$ in equations~\ref{eq:fresnel-u} is replaced by $h+h_0$. This correction compensates for the systematic error in  determination the mask plane position when focusing on the mask edges. 
No other fitting parameters were used in the model and the intensity distribution for all values of $h$ was simulated using the same $h_0$ parameter value.
The data demonstrate qualitative and semi-quantitative with the predictions of the approximate diffraction theory. This experiment also allows one to observe the gradual transition from the Fresnel to the Fraunhofer diffraction regime as the microscope objective moves away from the mask plane.

\subsection{Fraunhofer diffraction}
\label{FraunhoferSection}
\begin{figure*}[!htbp]
    \centering
    \includegraphics[width=\textwidth]{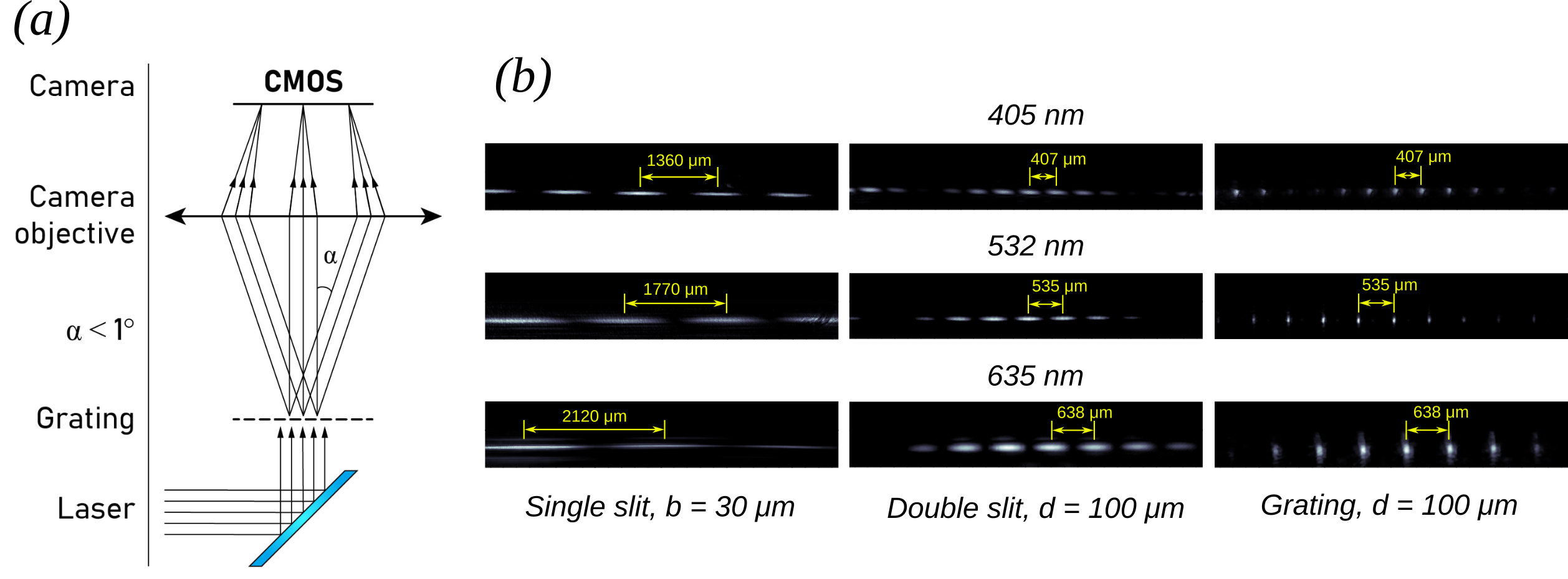}
    \caption{Fraunghofer diffraction patterns for bottom illumination with the laser pointers of different wavelengths ($405,\ 532$ and $635$\, nm). (a) Single slit $30\,\mu$m wide; (b) Double slit $100\,\mu$m apart; (c) Diffraction grating with period $100\,\mu$m.}
    \label{fig:fraunghofer}
\end{figure*}

 As described in previous section Fraunhofer diffraction occurs in the far-field regime, where only parallel beams form the pattern. As a result, its direct observation using the microscope is not a straightforward.
However, if one uses an empty slot in the nosepiece instead of infinity-corrected micro-objective, the diffracted light is focused by a tube lens directly onto CMOS matrix sensor, as shown in Fig.~\ref{fig:fraunghofer}a.
In order to observe diffraction pattern {within a sensor area} the diffraction angles must be small enough - in the range $\pm$2$^\circ$.
This is possible with 50 lines/mm or less diffraction gratings. Both top and bottom geometries are possible.
Light-emitting matrix of the LED display could be used instead of dedicated grating, as an easy available periodic structure.
In the bottom geometry it is also possible to observe diffraction on single slits with width above 20 $\mu$m, e.g. monochromator slits. Fig.~\ref{fig:fraunghofer}b shows diffraction patterns, obtained on different structures with bottom illumination with the laser pointers of three different wavelengths: $405,\ 532$ and $635$\, nm. 
We use laser pointer illumination because it is roughly parallel. LED light would require too small pinhole to be made similarly parallel that decreases the illumination of the detector.We show diffraction patterns from 
$30\,\mu$m wide single slit, double $30\,\mu$m slit with slit separation $100\,\mu$m and grating with period $100\,\mu$m.

The analysis of the obtained patterns is relatively straightforward since Fraunhofer diffraction by linear structures under small-angle diffraction conditions can be described through the Fourier transform of the transmission function. In particular, for a single slit the angular intensity distribution $I_1(\alpha)$ takes the form\cite{ghatak2009optics, hecht5ed}:
\begin{equation}
\label{eq:fraung-angular}
    I_1(\alpha) = \left[\dfrac{\sin u}{u}\right]^2, \quad u \approx \pi b \alpha/\lambda,
\end{equation}
where $\alpha$ is the diffraction angle, $b$ is the slit width, and $\lambda$ is the wavelength. Tube lens transforms 
the angular distribution into spatial one at the CMOS sensor plane:
\begin{equation}
\label{eq:fraung-linear}
    I_1(x) = \left[\dfrac{\sin v}{v}\right]^2, \quad v = \pi b x/(\lambda F),   
\end{equation}
where $F$ is the focal length of the tube lens, and $x$ is the linear distance from the center of the pattern to the observation point. The positions of the intensity minima can now be determined from the following condition:
\begin{equation}
\label{eq:fraung-minima}
    u = \pi m \quad \Rightarrow \quad x_m = \dfrac{\lambda}{b}F m,
\end{equation}
where $m$ is a non-zero integer. The positions of intensity maxima can be found as solutions to the transcendental equation $\tan u = u$.

For a double slit or diffraction grating, the intensity distribution formula becomes more complex due to interference between waves diffracted by different slits~\cite{ghatak2009optics, hecht5ed}:
\begin{equation}
\label{eq:fraung-double}
    I_N(x) = \left[\dfrac{\sin v}{v}\right]^2 \cdot \left[\dfrac{\sin(Nw)}{\sin w}\right]^2, \quad v = \dfrac{\pi b x}{\lambda F}, \quad w = \dfrac{\pi d x}{\lambda F},
\end{equation}
where $N$ is the total number of slits, $d$ is the separation between slits (grating period). In this case, the positions of principal maxima are given by:
\begin{equation}
\label{eq:fraung-minima2}
    x_m = \dfrac{\lambda}{d}F m.    
\end{equation}

Yellow numbers and arrows in Fig.~\ref{fig:fraunghofer}b {show the }
distances between the intensity maxima given by Eqs.\ref{eq:fraung-linear} and \ref{eq:fraung-minima2} with known geometrical parameters.
The observed patterns agree excellently with the diffraction theory.

\section{Discussion.}
\subsection{Parameters of the course}
Typically each exercise {in the} laboratory course includes understanding the theoretical background, performing the measurements, data analysis, error estimation, and report preparation. It takes from 1.5 to 3 hours in the lab to perform the exercise. For example, Newtons' rings could be observed with using lenses of various radii, at various wavelength, monochromaticity degree,{ and with} top or bottom illumination (see example of Instructional material in the Supplementary). The analysis could be either quantitative or qualitative.
Assuming a weekly laboratory schedule it takes several months to complete the full set of exercises and prepare reports. This laboratory course is supposed to run in parallel with lectures and problem solving sessions. 
The laboratory equipment appears to be rather cheap ($\sim$ 500 USD) and compact. This low entry threshold is crucial for newly established physical departments.
Although some basic optical topics, like goniometry, holography, total internal reflection, and common interferometers definitely require additional setups, the underlying physical laws and principles are covered by the microscope.

\subsection{Cost considerations}
\label{CostConsiderations}
{Table}~\ref{tab:components} shows a list of components that we bought using the Aliexpress platform. Similar or even better/cheaper components are available e.g. at Amazon or E-bay.

\begin{table}[t]
\caption{List of components for the 3D-printed microscope.}
\label{tab:components}
\centering
\begin{tabular}{c p{0.8\columnwidth} c}
\hline
\textbf{\#} & \textbf{Component and link} & \textbf{USD} \\
\hline
1  & Optical plate 200$\times$300\,mm (\href{https://aliexpress.com/item/1005006351786791.html}{Link}) & 100 \\
2  & Optical rod with mount; additional holes (6\,mm) drilled (\href{https://aliexpress.com/item/1005001296098768.html}{Link}) & 30 \\
3  & Coarse–fine vertical moving stage (\href{https://aliexpress.com/item/1005005903294013.html}{Link}) & 70 \\
4  & Microscope camera (C-mount) (\href{https://aliexpress.com/item/1005005440660706.html}{Link}) & 80 \\
5  & Achromatic doublet tube lens, $f\!\approx\!100$\,mm (\href{https://aliexpress.com/item/4000640392230.html}{Link}) & 10 \\
6  & Beamsplitter T/R 50/50 (\href{https://aliexpress.com/item/4000712297327.html}{Link}) & 10 \\
7  & Nosepiece for micro-objectives (\href{https://aliexpress.com/item/1005006396643185.html}{Link}) & 12 \\
8  & Infinity-corrected PLAN objectives 5$\times$ NA 0.13, 10$\times$ NA 0.25, 20$\times$ NA 0.40 (\href{https://aliexpress.com/item/1005003746358696.html}{Link}) & 170 \\
9  & LEDs for illumination (10 pcs) (\href{https://aliexpress.com/item/1005004740529823.html}{Link}) & 4 \\
10 & Plastic lens for illumination (\href{https://aliexpress.com/item/1005001832324314.html}{Link}) & 1 \\
11 & Interference filters 10$\times$10\,mm:
violet (\href{https://aliexpress.com/item/1005001367305454.html}{Link}), green (\href{https://aliexpress.com/item/4001256916589.html}{Link}), red (\href{https://aliexpress.com/item/4001257006358.html}{Link}) & 1 \\
12 & Iris diaphragm 1--12\,mm (\href{https://aliexpress.com/item/1005008797981242.html}{Link}) & 8 \\
\hline
   & \textbf{Total} & \textbf{$\sim$496} \\
\hline
\end{tabular}
\end{table}

{A second-hand microscope body (for $\sim$300 USD) could also be a good starting point instead of construction of the optic education system from scratch, taking several issues must be considered. Integrating custom illumination modules, filters, or beam splitters can be difficult in a commercial microscope. Photo-camera trinoculars assume much larger sensors than CMOS used here, resulting in a significant loss of the field of view. Leica and Zeiss objectives are not designed for the achromatic tube-lenses and are therefore not fully compatible with perfect infinity-corrected optics (like Nikon, Olympus or Mitutoyo). Another important issue is the optical thread compatibility. For teaching laboratories, 3D-printed microscopes offer a decisive advantage, as all instruments are identical by design and therefore ensure reproducible results across a group of students.}

Fabrication of c-mount threads requires a tap (\href{https://aliexpress.com/item/1005006949334577.html}{Link})/die (\href{https://aliexpress.com/item/2036103477.html}{Link}) couple (about 30 USD). 1 kilogram of plastic for 3D printer costs about 10 USD. Power source for LED Illumination could be done from USB charger with a few Ohm current-limiting resistor. The USB camera is connected to a PC. Alternatively, {the }camera can be connected to a display directly, but this option does not allow for image digitalization. Fraunhofer diffraction requires laser pointers, especially a violet one.

\begin{figure*}[!htbp]
\centering
\includegraphics[width=0.8\textwidth]{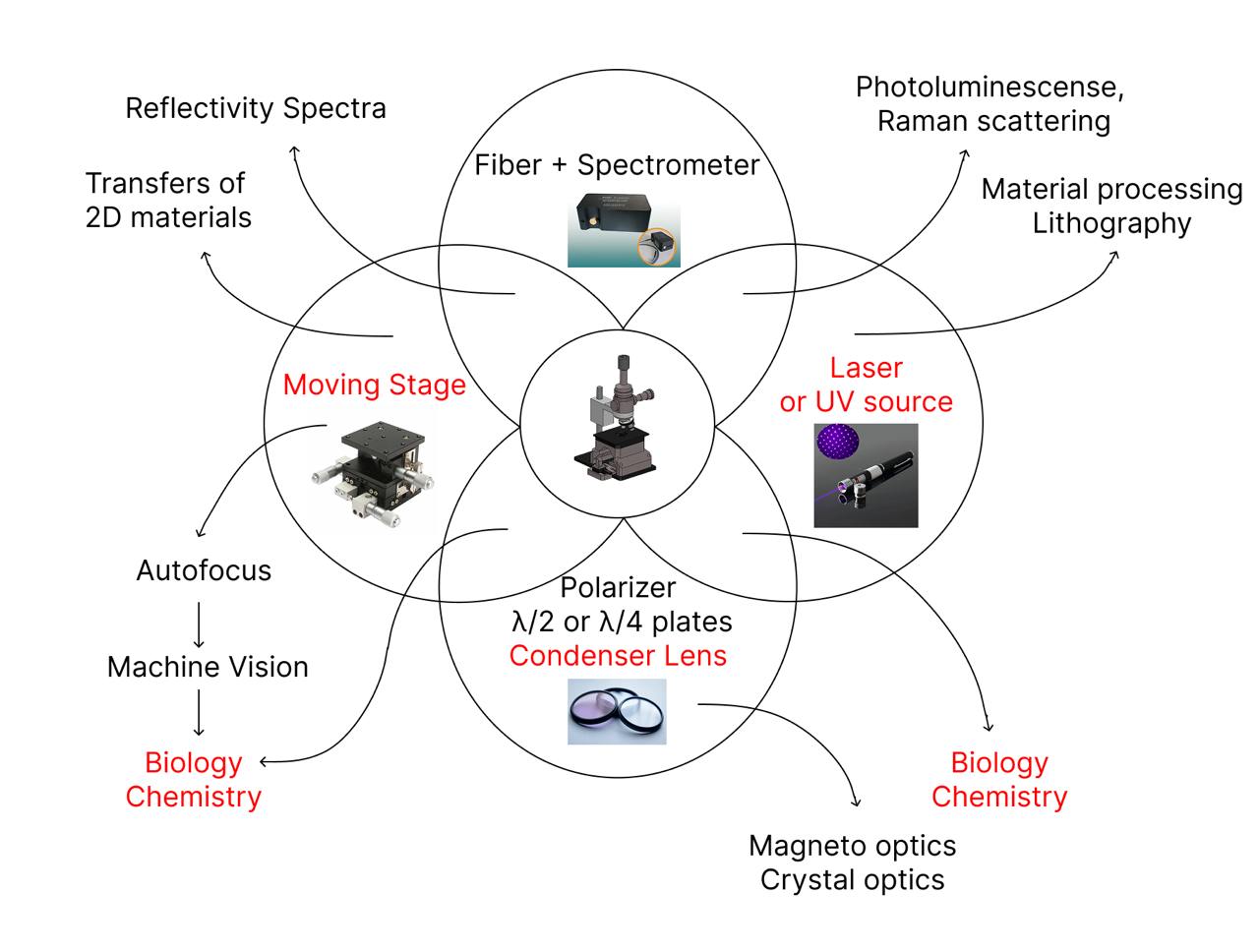}
\caption{Possible extensions of the microscope to the students research projects.}
\label{Extentions}
\end{figure*}

Thus the total components cost is about 500 USD. This is relatively cheap though the dimensions and availability of the parts may vary. 3D model of particular microscope must therefore be adjusted  to fit the component dimensions. Diffraction elements could be ordered at numerous photomask production websites and may cost a few hundred USD for 4 inch mask with the resolution of 2 $\mu$m. This is sufficient to produce at least 10 sets.

The microscope can be fabricated {even} cheaper. For example, beamsplitter could be replaced by a thin transparent plastic film, video-camera without a cage and C-mount thread costs about 15 USD, see \href{https://aliexpress.com/item/1005004404049549.html}{Link}). Instead of long working distance micro-objectives less sophisticated ones could be used, see e.g. two times cheaper PLAN objectives set \href{https://aliexpress.com/item/1005007239740128.html}{Link}). The basement could be 3D-printed and z-motion could be replaced by a standard linear micrometer-resolution stage \href{https://aliexpress.com/item/32848650820.htm}{Link}. The goal of our paper is not to demonstrate particular cost-effective solutions but rather to encourage researchers and teachers to build their own microscopy systems for education.

\subsection{Possible Extensions of the Microscope}

Additive manufacturing makes it easy to extend the capabilities of the microscope and apply it to various student's research projects. A scheme of possible directions is shown in Fig.\ref{Extentions}.
By installing an additional (possibly dichroic) beam splitter in the reflection geometry, a laser beam can be tightly focused into a spot using an infinity-corrected objective. This laser spot could be used for materials processing\cite{MoS2Topo}, especially so-called phase-change materials\cite{phasechange} and also in spectroscopy similarly to commercial Raman and photoluminescence setups\cite{CommerceConfocal}. An ultraviolet light source enables projection lithography\cite{Love,Galiullin} and also is very useful for visualizing the biological objects\cite{biolog}. A relatively inexpensive yet powerful extension is the reflectivity spectroscopy with a fiber spectrometer\cite{frisenda2017}, that will require a broadband intense parallel beam, an additional beam spitter, and fiber optic spectrometer input with a collimating lens that focuses light onto the fiber end.
{For} photoluminescence or Raman spectroscopy, it is essential to filter out the laser excitation. This can be achieved by replacing the 50/50 beam splitter with a dichroic mirror or a notch/edge optical filter.

Motorized control of the vertical position of either the microscope or the sample, combined with basic machine vision, enables automatic focusing. Motorized sample stage allows sample scanning. In combination with machine vision such scanning could be extremely useful for analysis of biological objects\cite{microfluidic}, or in field of 2D materials\cite{machinevision2D}. Within the undergraduate physics labs machine vision can also be applied to track Brownian motion of particles, an experiment commonly included in molecular physics curricula\cite{brownian}.
Noteworthy a field of 2D materials is becoming available for education process with microscope, see e.g. Ref.\cite{GrapheneUndergrad}. 3D printed microscope setup can be adapted for transferring 2D materials and assembling van der Waals heterostructures with the addition of simple mechanical manipulators \cite{Zhao2020, Martanov2020}.

The microscope capabilities could also be extended to crystal and magneto-optics by adding up polarization elements. By adding an optical arm, the microscope can be transformed into a functional interferometer\cite{InterferometerMicroscopeEducation}. Biological applications with high-NA objectives require a condenser lens for bottom illumination. 
All {these} extensions show {the great potential of home-made microscopy} and could easily serve as student's research projects those follow the optic educational course. Open hardware platforms, such as OpenFlexure\cite{OpenFlexure} and OpenFrame\cite{OpenFrame}, provide solutions for motorizing the system. Machine vision {integration with} deep learning algorithms\cite{DeepLearning} further enhances the microscope capabilities.

\section{Conclusion}

We demonstrate that essential part of the undergraduate physical optic practical exercises including interference and diffraction, could be covered with a home-made microscope. Almost monochromatic filtered and parallel LED illumination is required for observation of these phenomena that is produced by an illumination scheme including LED, two lenses, interference filter and a diphragm. Quantitative understanding of the wave optic-phenomena is possible through the digital image processing and proper calibration. The presented microscope allows further extensions to explore the other various scientific and engineering directions and encourages scientists and teachers to fabricate their microscope systems.

\section{Supplementary information}

Supplementary includes stl files of the 3D printed parts. Alternatively the files could be accessed via Ref. \href{https://cloud.mail.ru/public/JEpW/EYDKhV9q6}{Link}. Example of instructional materials on Newton's rings is also attached.

\begin{acknowledgments}

The work is supported by Russian Science Foundation grant 23-12-00340. Masks for diffraction were fabricated using Shared Facility Center of the P.N. Lebedev Physical Institute. The authors are thankful to G.V. Rybalchenko, Yu.G. Vainer and D.V. Kazantsev for discussions. 

\end{acknowledgments}

\end{document}